\documentclass[12pt]{article}

\newtheorem{lem}{Lemma}
\newtheorem{thm}[lem]{Theorem}

\newtheorem{define}{Definition}
\newtheorem{ex}{Example}
\newcommand{\qed}{\hfill $\Box$}
\newcommand{\proof}{\noindent {\bf Proof. } }

\usepackage{graphicx}
\usepackage{amsmath}
\usepackage{amsfonts}
\usepackage{latexsym}
\usepackage{amssymb}
\def\leg{\le}
\def\symdiff(#1,#2){#1 \Delta #2} %%% {(#1 \cup #2) - (#1 \cap #2)}
\long\def\longdelete#1{}

\title{{\scshape\normalsize Mathematics Division, National Center for
         Theoretical Sciences at Taipei}\\
         {\scshape\large NCTS/TPE-Math Technical Report 2004-013} \\~\\
\textbf{Diagnosabilities of Regular Networks}}
\author{Guey-Yun Chang\thanks{Department of Computer Science and
                              Information Engineering,
                              National Taiwan University,
                              Taipei 10617, Taiwan.}
                      \thanks{Email: d89006@csie.ntu.edu.tw.}
        \and
        Gerard J. Chang\thanks{Corresponding author. Department of Mathematics,
                               National Taiwan University,
                               Taipei 10617, Taiwan.
                               Member of Mathematics Division,
                               National Center for Theoretical Sciences at Taipei.
                               Email: gjchang@math. ntu.edu.tw.
                               Supported in part by the National Science
                               Council under grant NSC91-2213-E-002-128.}
                               \and
        Gen-Huey Chen$^*$\thanks{Email: ghchen@csie.ntu.edu.tw.}}
\date{October 29, 2003 (revision August 9, 2004)} %%% Gerard (title of sec 3)

\begin{document}
\maketitle

\begin{abstract}
\noindent In this paper, we study diagnosabilities of
multiprocessor systems under two diagnosis models: the PMC model
and the comparison model. In each model, we further consider two
different diagnosis strategies: the precise diagnosis strategy
proposed by Preparata {\it et al}.~and the pessimistic diagnosis
strategy proposed by Friedman. The main result of this paper is to
determine diagnosabilities of regular networks with certain
conditions, which include several widely used multiprocessor
systems such as variants of hypercubes and many others.
\end{abstract}


\newpage
\baselineskip 24pt

%**************************************************************************
%                                 Section 1
%*************************************************************************
\section{Introduction}

Fault diagnosis is an important step in the design of
multiprocessor systems and VLSI/WSI-oriented computing systems.
And automatic fault diagnosis has been considered
an integral part of the process of achieving fault tolerance.
A diagnosis strategy means a process to diagnose faults,
and it is {\it precise} (respectively, {\it pessimistic})
if no fault-free processor is mistaken as a faulty one
(respectively, a fault-free processor may be mistaken as a faulty one).
In order to diagnose faults,
a number of tests are performed among processors
and the collection of all test results is referred to as a {\it syndrome}.

Suppose that $S$ is a system with at most $t$ faulty processors.
Based on a precise diagnosis strategy,
$S$ is $t$-{\it diagnosable} if given any syndrome,
all faulty processors can be determined \cite{c2}.
The maximum $t$ for which $S$ is
$t$-diagnosable is called the {\it diagnosability} of $S$ \cite{c14}.
On the other hand,
based on a pessimistic diagnosis strategy,
$S$ is $t/s$-{\it diagnosable} if given any syndrome,
all faulty processors can be confined to a set of at most $s$ processors,
where $t\leq s$ \cite{c3}.
The maximum $t$ for which $S$ is $t$/$t$-diagnosable
is also called the {\it diagnosability} of $S$ \cite{c6}.

Preparata, Metzem, and Chien \cite{c2} first proposed a model,
called the {\it PMC model}, for fault diagnosis in a
multiprocessor system. Under the PMC model, all tests are
performed between two adjacent processors, and it was assumed that
a test result is reliable (respectively, unreliable) if the
processor that initiates the test is fault-free (respectively,
faulty). The PMC model was also adopted in \cite{c14}, \cite{c17},
\cite{c24}, \cite{c19}, \cite{c25}, \cite{c6} and \cite{c16}.

Malek \cite{c5} proposed another model, called the comparison
model, under which each test is initiated by a unique arbitrator.
The arbitrator feeds a pair of processors with the same task and
input and then compares their outputs. It is assumed that the
outputs are identical if they are fault-free, and distinct
otherwise. Only a fault-free arbitrator can guarantee a reliable
test result. Later, Maeng and Malek \cite{c7} modified Malek's
model so that multiple arbitrators were allowed and each
arbitrator can test any two of its adjacent processors. Maeng and
Malek's model is referred to as the {\it MM model}. Sengupta and
Dahbura \cite{c1} further suggested a modification of the MM
model, called the {\it MM* model}, in which any processor has to
test another two processors if the former is adjacent to the later
two. The MM* model was also adopted in \cite{c15}, \cite{c23} and
\cite{c18}.

Under the PMC model with a precise (respectively, pessimistic)
strategy, an $n$-dimensional hypercube has diagnosability $n$
\cite{c14} (respectively, $2n-2$ \cite{c6}); an $n$-dimensional
enhanced hypercube has diagnosability $n+1$ (respectively, 2$n$)
\cite{c16}; an $n$-dimensional M\"{o}bius cube has diagnosability
$n$ (respectively, $2n-2$) \cite{c17}; an $n$-dimensional star
graph has diagnosability $n-1$ (respectively, $2n-4$) \cite{c25}.
On the other hand, under the MM* model with a precise strategy, an
$n$-dimensional hypercube has diagnosability $n$ \cite{c18}; an
$n$-dimensional enhanced hypercube has diagnosability $n+1$
\cite{c18}; an $n$-dimensional crossed cube has diagnosability $n$
\cite{c23}; a $k$-ary $n$-dimensional butterfly graph has
diagnosability $2k-2$ if $k \geq 3$ and $n \geq 3$ \cite{c15}.

In this paper, we establish sufficient conditions for computing
diagnosabilities of regular networks. Our results are valid for
both the PMC and the MM* models with both the precise and the
pessimistic strategies. As consequences, diagnosabilities of many
well-known and unknown but potentially useful multiprocessor
systems can be obtained. These include hypercubes, enhanced
hypercubes, twisted cubes, crossed cubes, M\"{o}bius cubes,
cube-connected cycles, tori, star graphs, and many others. Some of
these are established in several papers as described in the
previous paragraph, and many are new.

In the next section, we introduce definitions and notations which
are used throughout this paper. We then derive in Section 3 the
diagnosabilities of regular networks with certain conditions under
different models and strategies. Consequently, the
diagnosabilities of several widely used multiprocessor systems are
determined in Section 4. Finally, in Section 5, we conclude the
paper with some remarks.

%************************************************************************
%                                Section 2
%************************************************************************
\section{Preliminaries}

In the study of multiprocessor systems,
the topology of a system is often adequately represented by a graph $G=(V,E)$,
where each node $u \in V$ denotes a processor and
each edge $(u,v) \in E$ denotes a link between nodes $u$ and $v$.
Previously, when the PMC model was adopted,
a self-diagnosable system was often represented by a directed graph
in which an arc directed from node $u$ to node $v$ means that $u$ can test $v$.
On the other hand, when the MM* model was adopted,
a self-diagnosable system was often represented by a multigraph
in which an edge $(u, v)$ labeled with $w$ means that $w$
is an arbitrator for $u$ and $v$, i.e.,
$w$ can test both $u$ and $v$. Since multiple
arbitrators for the same pair of nodes are allowed,
the representing graph can be a multigraph.

Throughout this paper we use a graph $G=(V, E)$ to represent a
self-diagnosable system. For a node $u$ of $G$, denote by $N(u)$
the set of all its neighboring nodes, i.e., $N(u)=\{v\in V: v$ is
adjacent to $u\}$. For a subset $S$ of $V$, let $N(S)=\cup_{v\in
S} N(v)$.

%************************** definition 1  ************************************
\begin{define}
 Under the PMC model, a syndrome $\sigma$ for system $G$ is defined as follows.
 For any two distinct nodes $u$ and $v$ with $v\in N(u)$,
$$
   \sigma (u,v)=\left\{\begin{array}[c]{ll}
                   0,  & \mbox{if $v$ is tested by $u$ to be fault-free}; \\
                   1,  & \mbox{if $v$ is tested by $u$ to be faulty}.
                         \end{array} \right.
$$
\end{define}

%************************** definition 2  ************************************
\begin{define}
   Under the MM* model, a syndrome $\sigma$ for system
   $G$ is defined as follows.
   For any three distinct nodes $u$, $v$ and $w$ with
   $u,v \in N(w)$,
 $$
   \sigma (u,v;w)=\left\{\begin{array}[c]{ll}
        0, & \mbox{if the test results of $u$ and $v$ by $w$ are identical}; \\
        1, & \mbox{if the test results of $u$ and $v$ by $w$ are distinct}.
                         \end{array} \right.
$$
\end{define}

\bigskip

Notice that the test result initiated by a faulty processor is
unreliable, and more than one syndrome may be produced for $G$
with faulty nodes. For each subset $F \subseteq V$, let
$\Omega(F)$ represent the set of syndromes that can be produced if
$F$ is the set of all faulty nodes. When $G$ has faulty nodes, a
syndrome $\sigma$ is randomly generated for the purpose of fault
diagnosis. We call $F$ an {\it allowable fault set with respect to
$\sigma$} under the PMC model (respectively, the MM* model) if (1)
and (2) hold (respectively, ($1^*$) and ($2^*$) hold).

(1)  $\sigma(u,v)=0$  for $u \in V-F$ and $v \in V-F$.

(2)  $\sigma(u,v)=1$  for $u \in V-F$ and $v \in F$.

($1^*$)  $\sigma(u,v;w)=0$ for $u \in V-F$, $v \in V-F$ and $w \in V-F$.

($2^*$)  $\sigma(u,v;w)=1$ for $(u \in F$ or $v \in F)$ and $w \in V-F$.

It is easy to see that $F$ is an allowable fault set with respect to $\sigma$
if and only if $\sigma \in \Omega(F)$.
Also, the set of all faulty nodes in $G$ is an
allowable fault set with respect to $\sigma$.

Two subsets $F_1$ and $F_2$ of $V$ are {\it distinguishable} if
$\Omega(F_1) \cap \Omega(F_2)=\emptyset$, and {\it
indistinguishable} otherwise. When $F_1$ and $F_2$ are
distinguishable, for each syndrome $\sigma$ in $\Omega(F_1) \cup
\Omega(F_2)$, exactly one of $F_1$ and $F_2$ is an allowable fault
set with respect to $\sigma$.  In this case, $F_1$ and $F_2$ are
distinct.  On the other hand, when $F_1$ and $F_2$ are
indistinguishable, they are allowable fault sets with respect to
each syndrome in $\Omega(F_1) \cap \Omega(F_2)$.

%************************** Definition 3 ************************************
\begin{define}          \label{def 3}
Under the precise diagnosis strategy, a system $G=(V,E)$ is
$t$-diagnosable if for any two subsets $F_1$ and $F_2$ of $V$ such
that $|F_1| \leg t $ and $|F_2| \leg t $, the sets $F_1$ and $F_2$
are distinguishable.
\end{define}

%************************** Definition 4 ************************************
\begin{define}          \label{def 4}
Under the pessimistic diagnosis strategy, a system $G=(V,E)$ is
$t/t$-diagnosable if for any two subsets $F_1$ and $F_2$ of $V$
such that $|F_1| \leg t$, $|F_2| \leg t$ and $|F_1\cup F_2|>t$,
the sets $F_1$ and $F_2$ are distinguishable.
\end{define}

\bigskip

The following characterization is useful for the
distinguishability of two sets under the MM* model. The {\it
symmetric difference} of two sets $A$ and $B$ is defined as the
set $A \Delta B = (A \cup B) - (A \cap B)$.

%**********************************       Lemma 4  ***********************
\begin{lem} {\bf (\cite{c1})}      \label{Lemma 4}
Suppose $G=(V,E)$ is a system under the MM* model. Two distinct
subsets $F_1$ and $F_2$ of $V$ are distinguishable if and only if
there is a node $v \in V-(F_1 \cup F_2)$ such that at least one of
the following conditions holds.

{\rm (1)} $|N(v)\cap(F_1- F_2)| \ge 2$.

{\rm (2)} $|N(v)\cap(F_2- F_1)| \ge 2$.

{\rm (3)} $|N(v)-(F_1 \cup F_2)| \ge 1$ and
          $|N(v)\cap (\symdiff(F_1,F_2))| \ge 1$.
\end{lem}

%************************************************************************
%                                Section 3
%************************************************************************
\section{Diagnosabilities of regular networks}

This section determines diagnosabilities of regular networks with
certain conditions.
% which include many well-know and unknown but
%potentially useful multiprocessor systems such as variants of
%hypercubes and many others.
Our results are for systems under the PMC model and the MM* model
each using both the precise and the pessimistic diagnosis
strategies.

%==============================================================
%     Section 3.1
%===============================================================
\subsection{Precise diagnosis strategy}\label{subsection 3.1}

%In this subsection, we consider results for regular networks under
%the PMC model and the MM* model each using  the precise diagnosis
%strategy. The main results are in Theorems \ref{Theorem 6} and
%\ref{Theorem 7}. To establish these results, we first prove a
%useful lemma, i.e. Lemma \ref{Lemma 5}, which is the base of both
%theorems.

A graph is called {\it r-regular} if every node in this graph has
the same degree $r$. A graph is {\it triangle-free} if it does not
contain a complete graph of three nodes as a subgraph. {\it All
networks in this subsection are $r$-regular and triangle-free such
that $N(u)\neq N(v)$ for every two adjacent nodes $u$ and $v$.}
With these conditions, we prove the $r$-diagnosability of networks
under the PMC model and the MM* model each using the precise
diagnosis strategy, see Theorems \ref{Theorem 6} and \ref{Theorem
7} respectively. Our plan is as follows.

Suppose to the contrary that $G$ is not $r$-diagnosable, in either
model.  Then, there are two indistinguishable and hence distinct
sets $F_1$ and $F_2$ with $|F_1| \le r$ and $|F_2| \le r$. Using
the conditions mentioned above for the networks, we first prove in
Lemma \ref{Lemma 5} that there is a node $w\in\symdiff(F_1,F_2)$
adjacent to some node $x\not\in F_1\cup F_2$.  (For the purpose of
discussion below, let $F_3$ denote the set of all such nodes $x$.)
This is mainly because the conditions on the networks force that
there are not too many edges between the nodes in $F_1\cup F_2$.
Having this lemma, the result for the PMC model then follows
easily from the definition. For the result under the MM* model, a
longer argument is needed.  By the aid of Lemma \ref{Lemma 4}
together with nodes in $F_3$, we first establish that $|F_1\cap
F_2|$ is as large as to be either $r-1$ or $r-2$. Consequently,
$F_1-F_2$ and $F_2-F_1$ both have at most two elements. These
restrict the shape of $G$ greatly. The rest of the proof is then
separated into two cases depending on the size of $F_1\cap F_2$.

We now start with the common lemma for the PMC model and the MM*
model.

%************************** Lemma 5  ************************************
\begin{lem}          \label{Lemma 5}
Suppose $r \ge 2$ and $G=(V,E)$ is an $r$-regular graph satisfying
the following two conditions.

{\rm (a)} $G$ is triangle-free.

{\rm (b)} $N(u)\neq N(v)$ for every two distinct nodes $u$ and $v$
of $G$.

\noindent Then, for any two distinct subsets $F_1$ and $F_2$ of
$V$ with $|F_1| \le r$ and $|F_2| \le r$, there exists a node $w
\in \symdiff(F_1,F_2)$ adjacent to some node $x \not\in F_1 \cup
F_2$.
\end{lem}

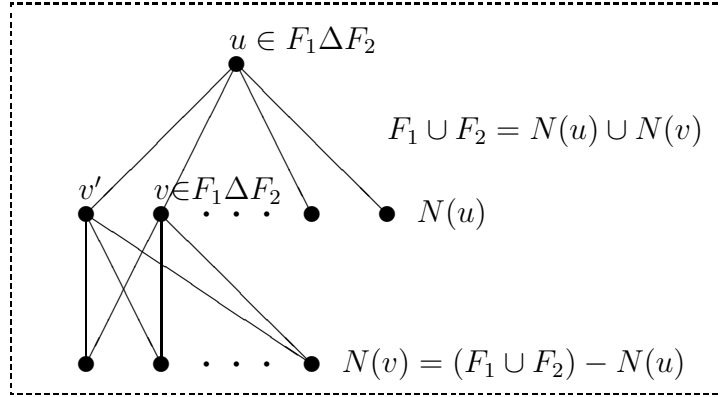
\begin{figure}[htb]
\setlength{\unitlength}{0.2cm}%{0.15cm}
\begin{center}
\begin{picture}(35,24)(1,10)
\put(10,30){\circle*{1.125}} \put(00,10){\circle*{1.125}}
\put(05,10){\circle*{1.125}} %\put(10,10){\circle*{1.125}}
\put(15,10){\circle*{1.125}} \put(00,20){\circle*{1.125}}
\put(05,20){\circle*{1.125}} \put(15,20){\circle*{1.125}}
\put(20,20){\circle*{1.125}} \put(20,25){$F_1\cup F_2=N(u)\cup
N(v)$} \put(9.5,31){$u\in\symdiff(F_1,F_2)$}
\put(4.5,21){$v$$\in$$\symdiff(F_1,F_2)$}\put(22,19.5){$N(u)$}
\put(17,9.5){$N(v)=(F_1\cup F_2)-N(u)$}\put(-0.5,21){$v'$}
\put(10,30){\line(-1,-1){10}} \put(10,30){\line(-1,-2){ 5}}
\put(10,30){\line( 1,-2){5}}
%\put(10,30){\line( 1,-2){ 5}}
\put(10,30){\line( 1,-1){10}}

\put(05,20){\line( -1, -2){5}} \put(05,20){\line( 0, -1){10}}
%\put(5,20){\line( 1, -2){5}}
\put(5,20){\line( 1, -1){10}}\put(00,20){\line( 0, -1){10}}
\put(00,20){\line( 1, -2){5}}
%\put(5,20){\line( 1, -2){5}}
\put(00,20){\line( 3, -2){15}}

\multiput(8,20)(2,0){3}{\circle*{0.4}}
\multiput(8,10)(2,0){3}{\circle*{0.4}}

 \put(-5,8){\dashbox{0.3}(48,26)}
\end{picture}
\end{center}
\caption{Relation among the sets in the proof of Lemma \ref{Lemma
5}.} \label{Figure 0}
\end{figure}

\proof Suppose to the contrary that $N(w) \subseteq F_1 \cup F_2$
for all nodes $w \in \symdiff(F_1,F_2)$. As $F_1 \ne F_2$, we may
choose $u \in \symdiff(F_1,F_2)$. In this case, $N(u) \subseteq
F_1 \cup F_2$. By the facts that $|N(u)|=r$ and $|F_1 \cap F_2| <
\max\{|F_1|, |F_2|\} \le r$, we know that $u$ has a neighbor $v
\in \symdiff(F_1,F_2)$. Again, we have $N(v) \subseteq F_1 \cup
F_2$. Since $G$ is triangle-free, $N(u) \cap N(v) =\emptyset$.
Therefore,
$$
   2r = |N(u)| + |N(v)| = |N(u) \cup N(v)| \le
   |F_1 \cup F_2| = |F_1|+|F_2|-|F_1 \cap F_2| \le 2r.
$$
Consequently, all inequalities are equalities and so $F_1\cap
F_2=\emptyset$ and $F_1\cup F_2$ is the disjoint union of $N(u)$
and $N(v)$.  Therefore, $N(v)=(F_1 \cup F_2) - N(u)$.  As $r \ge
2$, node $u$ has another neighbor $v'\ne v$. Since $F_1 \cap F_2
=\emptyset$, we have $v' \in \symdiff(F_1,F_2)$. By a similar
argument as above, we have $N(v')=(F_1 \cup F_2) - N(u)$ and so
$N(v)=N(v')$, a contradiction to condition (b). For the relation
among these sets, see Figure \ref{Figure 0}. \qed

\bigskip

According to Lemma \ref{Lemma 5} and the definition of
diagnosability of a system under the PMC model using the precise
diagnosis strategy, we have

%***************   Theorem 6 The PMC model by precise diagnosis strategy****
\begin{thm} \label{Theorem 6}
If $r\ge 2$ and $G$ is an $r$-regular graph, then $G$ is
$r$-diagnosable under the PMC model using the precise diagnosis
strategy if the following two conditions hold.

{\rm (a)} $G$ is triangle-free.

{\rm (b)} $N(u)\neq N(v)$ for every two distinct nodes $u$ and $v$
of $G$.
\end{thm}

\proof Suppose to the contrary that $G$ is not $r$-diagnosable.
Then, by Definition \ref{def 3}, there exist two indistinguishable
and hence distinct sets $F_1$ and $F_2$ with $|F_1| \le r$ and
$|F_2| \le r$. By Lemma \ref{Lemma 5}, there exists a node $w \in
\symdiff(F_1,F_2)$ adjacent to some node $x \not\in F_1 \cup F_2$.
Without loss of generality, we may assume that $w \in F_1 -F_2$.
Choose a syndrome $\sigma \in \Omega(F_1) \cap \Omega(F_2)$. If
$\sigma(x, w)=0$ (respectively, $\sigma(x,w)=1$), then $F_1$
(respectively, $F_2$) is not an allowable fault set with respect
to $\sigma$, a contradiction. \qed

\bigskip

For the discussion of the diagnosability under the MM$^*$ model
using the precise diagnosis strategy, we need the aid of Lemma
\ref{Lemma 5} as well as Lemma \ref{Lemma 4}. The result is
similar to that for the PMC model, except now there are two
exceptional networks defined as follows.

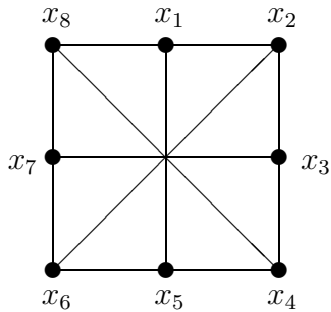
\begin{figure}[htb]
\setlength{\unitlength}{0.15cm}
\begin{center}
\begin{picture}(41,25)(0,10)
\put(09,07){$x_6$}
\put(19,07){$x_5$}
\put(29,07){$x_4$}
\put(06,19){$x_7$}
\put(32,19){$x_3$}
\put(09,32){$x_8$}
\put(19,32){$x_1$}
\put(29,32){$x_2$}
\multiput(10,10)(10,0){3}{\circle*{1.5}}
\multiput(10,20)(20,0){2}{\circle*{1.5}}
\multiput(10,30)(10,0){3}{\circle*{1.5}}
\put(10,10){\line(1,0){20}}
\put(10,20){\line(1,0){20}}
\put(10,30){\line(1,0){20}}
\put(10,10){\line(0,1){20}}
\put(20,10){\line(0,1){20}}
\put(30,10){\line(0,1){20}}
\put(10,10){\line(1,1){20}}
\put(10,30){\line(1,-1){20}}
\end{picture}
\end{center}
\caption{The graph $G_8$.}
\label{Figure 1}
\end{figure}

The first graph is $G_8$ obtained from a 8-cycle joining the 4
pairs of the farest vertices. More precisely, $G_8$ is the graph
with vertex set $V(G_8) = \{x_1, x_2, \ldots,x_8\}$ and edge set
$$E(G_8)= \{(x_i, x_{i+1}): 1\leq i\leq 7\} \cup
         \{(x_8,x_1)\} \cup
         \{(x_j, x_{j+4}): 1\leq j\leq 4\}.$$
See Figure \ref{Figure 1} for the graph $G_8$.

%%%%%%%%%%%%%%%%%%%

%\newpage

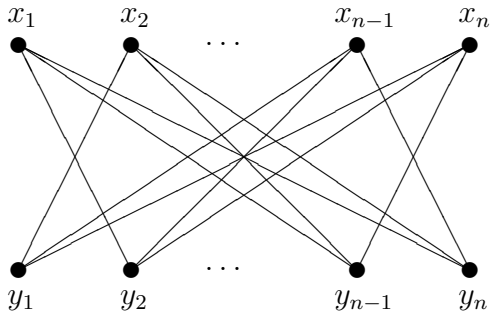
\begin{figure}[htb]
\setlength{\unitlength}{0.15cm}
\begin{center}
\begin{picture}(62,25)(0,10)
\put(09,07){$y_1$}
\put(19,07){$y_2$}
%\put(26.5,07){$\ldots$$\ldots$}
\put(38,07){$y_{n-1}$}
\put(49,07){$y_n$}
\put(09,32){$x_1$}
\put(19,32){$x_2$}
%\put(26.5,32){$\ldots$$\ldots$}
\put(38,32){$x_{n-1}$}
\put(49,32){$x_n$}
\put(10,10){\circle*{1.5}}
\put(20,10){\circle*{1.5}}
\put(26.5,10){$\ldots$}
\put(40,10){\circle*{1.5}}
\put(50,10){\circle*{1.5}}
\put(10,30){\circle*{1.5}}
\put(20,30){\circle*{1.5}}
\put(26.5,30){$\ldots$}
\put(40,30){\circle*{1.5}}
\put(50,30){\circle*{1.5}}
\put(10,10){\line(1,2){10}}
\put(10,10){\line(3,2){30}}
\put(10,10){\line(2,1){40}}
\put(20,10){\line(-1,2){10}}
\put(20,10){\line(1,1){20}}
\put(20,10){\line(3,2){30}}
\put(40,10){\line(-3,2){30}}
\put(40,10){\line(-1,1){20}}
\put(40,10){\line(1,2){10}}
\put(50,10){\line(-2,1){40}}
\put(50,10){\line(-3,2){30}}
\put(50,10){\line(-1,2){10}}
\end{picture}
\end{center}
\caption{The graph $G_{n,n}$.}
\label{Figure 2}
\end{figure}

The second graph is $G_{n,n}$ obtained from the complete bipartite
graph $K_{n,n}$ by removing a perfect matching. More formally,
$G(n,n)$ is the graph with vertex set $V(G_{n,n}) = \{x_1,
x_2,\ldots, x_n, y_1, y_2, \ldots,y_n\}$ and edge set
$$E(G_{n,n}) = \{(x_i, y_j): 1\leq i\leq n, 1\leq j\leq n \mbox{
and } i\neq j\}.$$ See Figure \ref{Figure 2} for the graph
$G_{n,n}$.

%%%%%%%%%%%%%%%%%%%

We are now ready to establish diagnosabilities for regular
networks under MM* model using the precise diagnosis
strategy.% by apagoge.
%First, assume the contrary and prove there
%exists $x \not\in F_1 \cup F_2$ such that  $|N(x) \cap (F_1\cap
%F_2)| \ge r-2$ and $N(x) \subseteq F_1\cup F_2$.  That is, the
%value of $|F_1\cap F_2|$ is $r-2$ or $r-1$. Note that $|F_1 \cup
%F_2- N(x)|=|F_1|+|F_2|-|F_1\cap F_2|-|N(x)|\le r+r-(r-2)-r=2$.
%Second, denote $F_3$ the set of all such $x$. Prove $|N(F_3)| >
%|F_1\cup F_2|$ which is a contradiction.

%*************     Theorem 7  The MM* model by precise diagnosis strategy******
\begin{thm} \label{Theorem 7}
If $r \ge 3$ and $G$ is an $r$-regular graph, which is not
isomorphic to $G_8$ or $G_{r+1,r+1}$, then $G$ is $r$-diagnosable
under the MM* model using the precise diagnosis strategy if the
following two conditions hold.

{\rm (a)} $G$ is triangle-free.

{\rm (b)} $N(u)\neq N(v)$ for every two distinct nodes $u$ and $v$
of $G$.
\end{thm}

\proof  Suppose to the contrary that $G$ is not $r$-diagnosable.
Then, by Definition \ref{def 3}, there exist two indistinguishable
and hence distinct sets $F_1$ and $F_2$ with $|F_1| \le r$ and
$|F_2| \le r$.  According to Lemma \ref{Lemma 5},
$\symdiff(F_1,F_2)$ has at least one node $w$ adjacent to some
node $x \not\in F_1\cup F_2$. Denote $F_3$ the set of all such
nodes $x$. Since $F_1$ and $F_2$ are indistinguishable, none of
the conditions in Lemma \ref{Lemma 4} holds. It follows that for
any node $v \in F_3$, we have

\medskip
 (i) $|N(v)\cap(F_1-F_2)|\leq 1$,

 (ii) $|N(v)\cap(F_2-F_1)|\leq 1$,

 (iii) $N(v) \subseteq F_1 \cup F_2$.

\medskip \noindent
Then, $F_3$ is an independent set with $N(F_3)
\subseteq F_1 \cup F_2$. Also, (i) and (ii) and $|N(v)|=r$ imply
$|N(v) \cap F_1 \cap F_2| \ge r-2$, which gives $|F_1 \cap F_2|
\ge r-2$ and so $|F_1 \cap F_2|=r-1$ or $r-2$.

Choose a node $w\in\symdiff(F_1,F_2)$ which is adjacent to some
node $x\in F_3$. Then, $
    |F_1 \cup F_2| \ge |N(x)| \ge r.
$ Suppose $|F_1 \cup F_2|=r$.   By (iii), $N(v)=F_1\cup F_2$ for
all nodes $v\in F_3$. Condition (b) then implies that $F_3$ has
just one node, which is $x$. In this case, $w$ must be adjacent to
all other nodes in $F\cup F_2$.  Thus a triangle forms, a
contradiction. Hence, $|F_1 \cup F_2| \ge r+1$.

Let $F_3 = \{v_1, v_2, \ldots, v_s\}$ and consider the following
two cases.

{\bf Case 1.} $|F_1 \cap F_2|=r-1$. In this case, $|F_1 \cup
F_2|=r+1$ and $|\symdiff(F_1,F_2)|=2$.

Let $F_1 \cup F_2 = \{w_1, w_2, \ldots, w_{r+1} \}$. As $G$ is
$r$-regular, (iii) and condition (b) imply $s \le r+1$ and,
without loss of generality, $N(v_i) = (F_1 \cup F_2)-\{w_i\}$ for
$1 \le i \le s$. We claim that $F_1 \cup F_2$ is independent.
Suppose to the contrary that $w_j$ is adjacent to $w_k$ for some
$j<k$. Since $G$ is triangle-free, any two neighbors of node $v_i$
are not adjacent. Hence, $w_j w_k \in E$ implies that $j=s=1$ or
$j=1 <k=s=2$. As $|\symdiff(F_1,F_2)|=2$, we may choose a vertex
$w_i\ne w_1$ from $\symdiff(F_1,F_2)$. Then $N(w_i) \subseteq
\{w_1\}\cup F_3$ and so $w_i$ has degree at most $1+s\le 3$ and
hence exactly $3$. Furthermore, $s=2$ and $w_i=w_2$, which is
adjacent to $v_1$ and $v_2$, contradicting that $v_2$ is not
adjacent to $w_2$.  So, $F_1\cup F_2$ is an independent set. In
this case, $N(w_p) \subseteq F_3$ and $N(w_q) \subseteq F_3$ for
the two nodes $w_p, w_q \in \symdiff(F_1,F_2)$. Condition (b) and
$s\le r+1$ then imply that $s=r+1$ and so $G \cong G_{r+1,r+1}$,
which is impossible.

{\bf Case 2.} $|F_1 \cap F_2|=r-2$. In this case, $|F_1-F_2| \le
2$ and $|F_2-F_1| \le 2$.

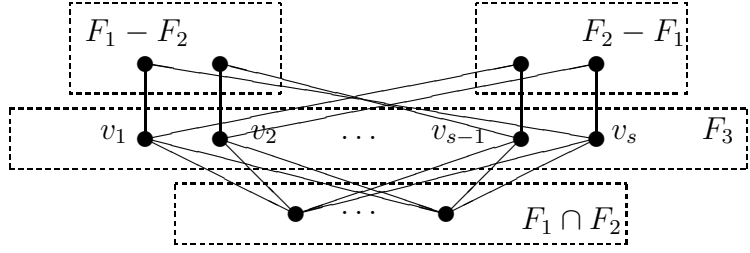
\begin{figure}[htb]
\setlength{\unitlength}{0.2cm}%{0.15cm}
\begin{center}
\begin{picture}(35,20)(0,5)
\put(-4,16.5){$F_1-F_2$} \put(29,16.5){$F_2-F_1$}
\put(37,10){$F_3$}\put(25,4){$F_1\cap F_2$}

 \put(-3,10){$v_1$}
\put(7,10){$v_2$}

\put(19,10){$v_{s-1}$} \put(31,10){$v_s$}

\put(0,10){\circle*{1.125}} \put(5,10){\circle*{1.125}}

\put(13,10){$\ldots$}

 \put(25,10){\circle*{1.125}}
\put(30,10){\circle*{1.125}}
 \put(0,15){\circle*{1.125}}
\put(5,15){\circle*{1.125}} \put(25,15){\circle*{1.125}}
\put(30,15){\circle*{1.125}}
 \put(10,5){\circle*{1.125}}
\put(13,5){$\ldots$} \put(20,5){\circle*{1.125}}

\put(0,10){\line(0,1){5}} \put(0,10){\line(5,1){25}}
\put(0,10){\line(4,-1){20}} \put(0,10){\line(2,-1){10}}
\put(5,10){\line(0,1){5}} \put(5,10){\line(1,-1){5}}
\put(5,10){\line(3,-1){15}} \put(5,10){\line(5,1){25}}

\put(25,10){\line(0,1){5}} \put(25,10){\line(-3,-1){15}}
\put(25,10){\line(-4,1){20}}\put(25,10){\line(-1,-1){5}}
 \put(30,10){\line(0,1){5}}
\put(30,10){\line(-6,1){30}} \put(30,10){\line(-2,-1){10}}
\put(30,10){\line(-4,-1){20}}\put(30,10){\line(0,1){5}}
 \put(-9,8){\dashbox{0.3}(49,4)}
\put(-5,13){\dashbox{0.3}(14,6)} \put(22,13){\dashbox{0.3}(14,6)}
\put(2,3){\dashbox{0.3}(30,4)}
\end{picture}
\end{center}
\caption{$|F_1-F_2| \le 2$ and $|F_2-F_1| \le 2$ imply $s\le 4$.}
\label{Figure 3}
\end{figure}

By (i)--(iii), $N(v_i)=(F_1\cap F_2) \cup\{v'_i,v''_i\}$ for each
$v_i \in F_3$, where $v'_i\in F_1-F_2$ and $v''_i\in F_2-F_1$.
Notice that the nodes $v'_i$ (respectively, $v''_i$) are not
necessarily distinct, but the sets $\{v'_i,v''_i\}$ are distinct.
Then, $|F_1-F_2| \le 2$ and $|F_2-F_1| \le 2$ imply $s\le 4$. For
the relation among these sets, see Figure \ref{Figure 3}.

Since $G$ is triangle-free, neighbors of $v'_i$ and $v''_i$ are in
$F_3$ or in $(\symdiff(F_1,F_2))-\{v'_i,v''_i\}$. We first give
four observations.

(1) If $N(v'_i) \cap F_3 = \{v_i\}$, then the other neighbors of
$v'_i$ are in $(\symdiff(F_1,F_2))-\{v'_i,v''_i\}$, which has at
most two nodes.  Hence, $N(v'_i)=\{v_i\} \cup
((\symdiff(F_1,F_2))-\{v'_i,v''_i\})$ has exactly $3$ nodes and
$r=3$ and $|\symdiff(F_1,F_2)|=4$.

(2) If $N(v'_i) \cap F_3 = \{v_i,v_j\}$ with $i\ne j$, then the
other neighbors of $v'_i$ are in
$(\symdiff(F_1,F_2))-\{v'_i,v''_i,v''_j\}$, which has at most one
node. Hence, $N(v'_i)=\{v_i,v_j\} \cup
((\symdiff(F_1,F_2))-\{v'_i,v''_i,v''_j\})$ has exactly $3$ nodes
and $r=3$ and $|\symdiff(F_1,F_2)|=4$.

(3) If there are at least $3$ distinct nodes  $v_i, v_j, v_k \in
N(v'_i) \cap F_3$, then $F_2-F_1$ contains at least three distinct
nodes $v''_i, v''_j, v''_k$, which is impossible.

(4) Similarly, either $N(v''_i)=\{v_i\} \cup
((\symdiff(F_1,F_2))-\{v'_i,v''_i\})$ or $N(v''_i)=\{v_i,v_j\}
\cup ((\symdiff(F_1,F_2))-\{v'_i,v'_j,v''_i\})$. In either case,
$N(v''_i)$ has exactly $r=3$ nodes and $|\symdiff(F_1,F_2)|=4$.

Having the four observations, we now continue our proof. If
$|N(v'_i)\cap F_3|=|N(v''_i)\cap F_3|=1$ for some $i$, then
$N(v'_i)=\{v_i\} \cup ((\symdiff(F_1, F_2)) - \{v'_i, v''_i\}) =
N(v''_i)$ by (1) and (4), contradicts condition (b).

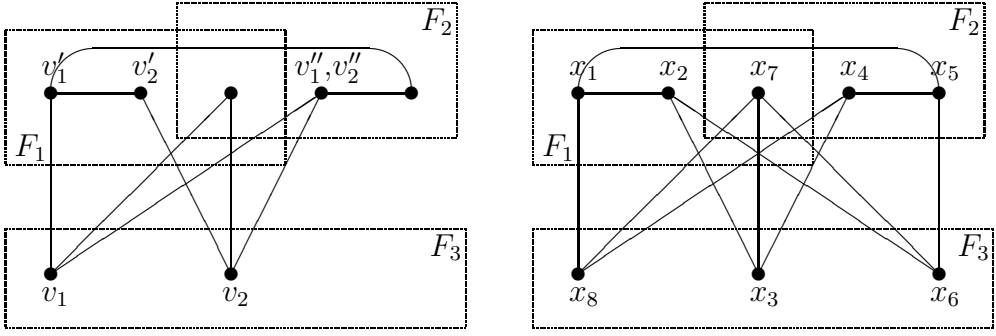
\begin{figure}[htb]
\setlength{\unitlength}{0.12cm}
\begin{center}
\mbox{
\begin{picture}(55,35)(03,6)
\put(09,07){$v_1$} \put(29,07){$v_2$} %\put(49,07){$x_6$}
\put(09,32){$v'_1$} \put(19,32){$v'_2$} %\put(29,32){$x_7$}
\put(37,32){$v''_1$,$v''_2$} %\put(49,32){$x_5$}
\put(10,10){\circle*{1.5}} \put(30,10){\circle*{1.5}}
%\put(50,10){\circle*{1.5}}
\put(10,30){\circle*{1.5}}
\put(20,30){\circle*{1.5}} \put(30,30){\circle*{1.5}}
\put(40,30){\circle*{1.5}} \put(50,30){\circle*{1.5}}
\put(10,10){\line(0,1){20}} \put(10,10){\line(1,1){20}}
\put(10,10){\line(3,2){30}} \put(30,10){\line(-1,2){10}}
\put(30,10){\line(0,1){20}} \put(30,10){\line(1,2){10}}
%\put(50,10){\line(-3,2){30}} \put(50,10){\line(-1,1){20}}
%\put(50,10){\line(0,1){20}}
\put(10,30){\line(1,0){10}}
\put(40,30){\line(1,0){10}} \put(30,30){\oval(40,10)[t]}

\put(5,4){\dashbox{0.2}(51,11)}
  %\multiput(5,4)(2,0){26}{\line(1,0){1}}
  %\multiput(5,15)(2,0){26}{\line(1,0){1}}
  %\multiput(5,4)(0,2){6}{\line(0,1){1}}
  %\multiput(56,4)(0,2){6}{\line(0,1){1}}
\put(5,22){\dashbox{0.2}(31,15)}
  %\multiput(5,22)(2,0){16}{\line(1,0){1}}
  %\multiput(5,37)(2,0){16}{\line(1,0){1}}
  %\multiput(5,22)(0,2){8}{\line(0,1){1}}
  %\multiput(36,22)(0,2){8}{\line(0,1){1}}
\put(24,25){\dashbox{0.2}(31,15)}
  %\multiput(24,24)(2,0){16}{\line(1,0){1}}
  %\multiput(24,39)(2,0){16}{\line(1,0){1}}
  %\multiput(24,24)(0,2){8}{\line(0,1){1}}
  %\multiput(55,24)(0,2){8}{\line(0,1){1}}
\put(52,12){$F_3$} \put(6,23){$F_1$} \put(51,37){$F_2$}
\end{picture}
}
%------------------------------
\mbox{
\begin{picture}(55,35)(03,6) \put(09,07){$x_8$}
\put(29,07){$x_3$} \put(49,07){$x_6$} \put(09,32){$x_1$}
\put(19,32){$x_2$} \put(29,32){$x_7$} \put(39,32){$x_4$}
\put(49,32){$x_5$} \put(10,10){\circle*{1.5}}
\put(30,10){\circle*{1.5}} \put(50,10){\circle*{1.5}}
\put(10,30){\circle*{1.5}} \put(20,30){\circle*{1.5}}
\put(30,30){\circle*{1.5}} \put(40,30){\circle*{1.5}}
\put(50,30){\circle*{1.5}} \put(10,10){\line(0,1){20}}
\put(10,10){\line(1,1){20}} \put(10,10){\line(3,2){30}}
\put(30,10){\line(-1,2){10}} \put(30,10){\line(0,1){20}}
\put(30,10){\line(1,2){10}} \put(50,10){\line(-3,2){30}}
\put(50,10){\line(-1,1){20}} \put(50,10){\line(0,1){20}}
\put(10,30){\line(1,0){10}} \put(40,30){\line(1,0){10}}
\put(30,30){\oval(40,10)[t]}

\put(5,4){\dashbox{0.2}(51,11)}
  %\multiput(5,4)(2,0){26}{\line(1,0){1}}
  %\multiput(5,15)(2,0){26}{\line(1,0){1}}
  %\multiput(5,4)(0,2){6}{\line(0,1){1}}
  %\multiput(56,4)(0,2){6}{\line(0,1){1}}
\put(5,22){\dashbox{0.2}(31,15)}
  %\multiput(5,22)(2,0){16}{\line(1,0){1}}
  %\multiput(5,37)(2,0){16}{\line(1,0){1}}
  %\multiput(5,22)(0,2){8}{\line(0,1){1}}
  %\multiput(36,22)(0,2){8}{\line(0,1){1}}
\put(24,25){\dashbox{0.2}(31,15)}
  %\multiput(24,24)(2,0){16}{\line(1,0){1}}
  %\multiput(24,39)(2,0){16}{\line(1,0){1}}
  %\multiput(24,24)(0,2){8}{\line(0,1){1}}
  %\multiput(55,24)(0,2){8}{\line(0,1){1}}
\put(52,12){$F_3$} \put(6,23){$F_1$} \put(51,37){$F_2$}
\end{picture}
}
\end{center} \caption{$G$ is isomorphic to $G_8$.}
\label{Figure 4}
\end{figure}

Now, by symmetric, assume that $N(v'_1) \cap F_3 = \{v_1\}$ and
$N(v''_1) \cap F_3 = \{v_1,v_2\}$. By (1) and (4), the adjacency
of the related nodes are shown as in the left of Figure
\ref{Figure 4}. As $v'_2$ is of degree $3$, it must be adjacent to
one more node in $F_3$, say $v_3$.  This implies that $G$ is in
fact $G_8$ as in the right of Figure \ref{Figure 4}.

The case of $|N(v'_1) \cap F_3| = |N(v''_1) \cap F_3| = 2$ is
similar, except now $v'_1$ is $x_2$. \qed

%================================================================
%     Section 3.2
%=================================================================
\subsection{Pessimistic diagnosis strategy}\label{subsection 3.2}

In parallel to the results of last subsection, in this subsection
we establish $(2r-2)/(2r-2)$-diagnosability of networks under the
PCM model and the MM* model each using the pessimistic diagnosis
strategy, see Theorems \ref{Theorem 9} and \ref{Theorem 10}
respectively. Arguments here are slightly more complicated than
those in the previous subsection, and stronger conditions on the
networks are necessary. More precisely, {\it all networks
considered in this subsection are $r$-regular and triangle-free
such that $|N(u)\cap N(v)| \le 2$ for every two distinct nodes $u$
and $v$.} Notice that the condition $|N(u) \cap N(v)| \le 2$ is
stronger than that $N(u) \ne N(v)$.  In fact, when $G$ is
$r$-regular, the former implies $|N(u) \cup N(v)| \ge 2r-2$ while
the later only implies $|N(u) \cup N(v)| \ge r+1$. For technical
reason, we also have an exceptional graph $G_5$ which is the graph
with vertex set
 $ V_5 = \{z, z_1, z_2, z_3, z_4, z_5\} \cup
         \{z_I: I \subseteq \{1,2,3,4,5\}, |I|=2\}
 $
 and edge set
 $
 E_5 = \{z z_i, z_i z_I, z_I z_J: i \in \{1,2,3,4,5\}, i \in I,
                         I ,J\subseteq\{1,2,3,4,5\},
                         |I|=|J|=2, I\cap J=\emptyset\}.
 $

\begin{figure}[htb]
\setlength{\unitlength}{0.2cm}%{0.15cm}
\begin{center}
\begin{picture}(48,25)(00,6)
\put(00,10){\circle*{1.125}} \put(05,10){\circle*{1.125}}
\put(10,10){\circle*{1.125}} \put(15,10){\circle*{1.125}}
\put(20,10){\circle*{1.125}} \put(25,10){\circle*{1.125}}
\put(30,10){\circle*{1.125}} \put(35,10){\circle*{1.125}}
\put(40,10){\circle*{1.125}} \put(45,10){\circle*{1.125}}

\put(00.5,8.5){$z_{12}$} \put(05.5,8.5){$z_{34}$}
\put(10.5,8.5){$z_{15}$} \put(15.5,8.5){$z_{23}$}
\put(20.5,8.5){$z_{45}$} \put(25.5,8.5){$z_{13}$}
\put(30.5,8.5){$z_{25}$} \put(35.5,8.5){$z_{14}$}
\put(40.5,8.5){$z_{35}$} \put(45.5,8.5){$z_{24}$}

\put(10,20){\circle*{1.125}} \put(15,20){\circle*{1.125}}
\put(20,20){\circle*{1.125}} \put(25,20){\circle*{1.125}}
\put(30,20){\circle*{1.125}}

\put(11,20){$z_1$} \put(16,20){$z_2$} \put(21,20){$z_3$}
\put(26,20){$z_4$} \put(31,20){$z_5$}

%\put(25,20){\circle*{1.125}} \put(30,20){\circle*{1.125}}
%\put(35,20){\circle*{1.125}}

\put(20,30){\circle*{1.125}} \put(21.5,29.5){$z$}

\put(20,30){\line(-1,-2){5}} \put(20,30){\line(-1,-1){ 10}}
\put(20,30){\line( 0,-1){10}} \put(20,30){\line( 1,-1){ 10}}
\put(20,30){\line( 1,-2){5}}

 \put(00,10){\line( 45, 0){45}}
 \put(10,20){\line( -1, -1){10}}
\put(10,20){\line( 0, -1){10}} \put(10,20){\line( 3, -2){15}}
\put(10,20){\line(5, -2){25}} \put(15,20){\line( -3, -2){15}}
\put(15,20){\line( 0, -1){10}} \put(15,20){\line( 3, -2){15}}
\put(15,20){\line( 3, -1){30}} \put(20,20){\line( -3, -2){15}}
\put(20,20){\line(-1, -2){5}} \put(20,20){\line(1, -2){5}}
\put(20,20){\line(2, -1){20}} \put(25,20){\line(-2, -1){20}}
\put(25,20){\line(-1, -2){5}} \put(25,20){\line(1, -1){10}}
\put(25,20){\line(2, -1){20}} \put(30,20){\line(-2, -1){20}}
\put(30,20){\line(0, -1){10}} \put(30,20){\line(1, -1){10}}
\put(30,20){\line(-1, -1){10}}

\put(17.5,10){\oval(25,5)[b]}\put(25,10){\oval(20,7)[b]}
\put(10,10){\oval(20,8)[b]} \put(20,10){\oval(40,10)[b]}
\put(35,10){\oval(20,12)[b]}\put(27.5,10){\oval(35,14)[b]}
\end{picture}
\end{center}
\caption{The graph $G_5$, where $z_{ij}$ stands for
$z_{\{i,j\}}$.} \label{Figure 5}
\end{figure}
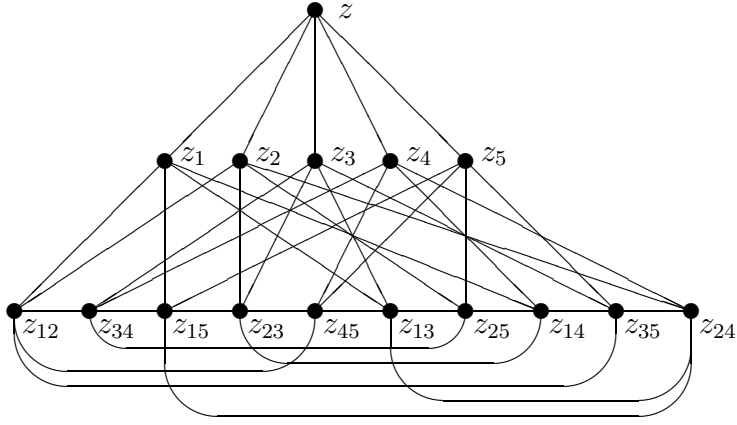

Our plan is as follows. Suppose to the contrary that $G$ is not
$(2r-2)/(2r-2)$-diagnosable, in either model.  Then, there are two
indistinguishable and hence distinct sets $F_1$ and $F_2$ with
$|F_1| \le 2r-2$ and $|F_2| \le 2r-2$ but $|F_1 \cup F_2| > 2r-2$.
Using the conditions mentioned above for the networks, we first
prove in Lemma \ref{Lemma 8} that there is a node
$w\in\symdiff(F_1,F_2)$ adjacent to some node $x\not\in F_1\cup
F_2$.  (For the purpose of discussion below, let $F_3$ denote the
set of all such nodes $x$.) Although the proof for Lemma
\ref{Lemma 8} is longer than that for Lemma \ref{Lemma 5}, the
main reason is also that the conditions on the networks force that
there are not too many edges between the nodes in $F_1\cup F_2$.
Having this lemma, again, the result for the PMC model follows
easily from the definition. For the result under the MM* model,
again, a longer argument is needed. By the aid of Lemma \ref{Lemma
4} together with nodes in $F_3$, we first establish that $|F_1\cap
F_2|\ge r-2$ and $|F_1\cup F_2| \le 3r-2$. It is then proved that
$|N(w) \cap F_3| \le 2$ for each node $w \in \symdiff(F_1,F_2)$.
These restrict the connections between $\symdiff(F_1,F_2)$ and
$F_3$. The rest of the proof is then separated into three cases
depending on the sizes of $F_3$ and $N(p)\cap (\symdiff(F_1,F_2))$
for $p\in F_3$.

We now start with the common lemma for the PMC model and the MM*
model.

\longdelete{%-----------------------
The following lemma proves there exists a node $w \in
\symdiff(F_1,F_2)$ adjacent to some node $x \not\in F_1 \cup F_2$
for any two distinct subsets $F_1$ and $F_2$ of $V$ with $|F_1|
\le 2r-2$ and $|F_2| \le 2r-2$ but $|F_1 \cup F_2|
> 2r-2$.  First, assume the contrary and prove there
exists $z \in F_1 \Delta F_2$ such that $N(z) \cap (F_1\Delta
F_2)=\{z_1,z_2,\ldots,z_s\}$ with $s\ge 2$. Note that $\{z\} \cup
N(z) \cup N(z_1)\cup \ldots \cup N(z_s)$ is a subset of $F_1 \cup
F_2$. In this limited case, prove $|F_1 \cup F_2| < |\{z\} \cup
N(z)\cup N(z_1)\cup \ldots N(z_s)|$, which is a contradiction.
 }%---------------------------------------------------------

%************************** Lemma 8  ************************************
\begin{lem}          \label{Lemma 8}
Suppose $r \ge 5$ and $G$ is an $r$-regular graph, which is not
isomorphic to $G_5$ and satisfies the following two conditions.

{\rm (a)} $G$ is triangle-free.

{\rm (b)} $|N(u) \cap N(v)| \le 2$ for every two distinct nodes
          $u$ and $v$ of $G$.

\noindent Then, for any two distinct subsets $F_1$ and $F_2$ of
$V$ with $|F_1| \le 2r-2$ and $|F_2| \le 2r-2$ but $|F_1 \cup F_2|
> 2r-2$, there exists a node $w \in \symdiff(F_1,F_2)$ adjacent to
some node $x \not\in F_1 \cup F_2$.
\end{lem}

\proof Suppose to the contrary that $N(w) \subseteq F_1 \cup F_2$
for all $w \in \symdiff(F_1,F_2)$. By the assumptions, $F_1 \cap
F_2$ is a proper subset of $F_1$ and $F_2$, and so
$|\symdiff(F_1,F_2)| \ge 2$. We may choose two distinct vertices
$u$ and $v$ from $\symdiff(F_1,F_2)$. If $N(u)$ and $N(v)$ are
subsets of $F_1 \cap F_2$, then condition (b) implies that
$$
   |F_1 \cap F_2| \ge |N(u)\cup N(v)| = |N(u)|+|N(v)|-|N(u)\cap N(v)|
   \ge r + r - 2 = 2r-2 \ge |F_1|,
$$
contradicting to that fact that $F_1 \cap F_2$ is a proper subset
of $F_1$.

Therefore, either $u$ or $v$ is adjacent to a vertex in $\symdiff(F_1,F_2)$.
So, we may choose two adjacent vertices $x$ and $y$ from $\symdiff(F_1,F_2)$.
If $N(x)-\{y\}$ and $N(y)-\{x\}$ are subsets of $F_1 \cap F_2$,
then condition (a) implies that $(N(x)-\{y\}) \cap (N(y)-\{x\}) = \emptyset$
and so
$$
   |F_1 \cap F_2| \ge |(N(x)-\{y\}) \cup (N(y)-\{x\})| =
   |N(x)-\{y\}|+|N(y)-\{x\}| =  (r-1)+(r-1)  \ge |F_1|,
$$
again a contradiction.

This proves that $\symdiff(F_1,F_2)$ has a vertex adjacent to at
least two vertices in $\symdiff(F_1,F_2)$. Now, choose a vertex $z
\in \symdiff(F_1,F_2)$ with a maximum number $s$ of neighbors in
$\symdiff(F_1,F_2)$, where $2 \le s\le r$. Let these $s$ neighbors
of $z$ be $z_1, z_2, \ldots, z_s$, and $A =\cup_{1\le i\le s}
(N(z_i)-\{z\})$. By condition (a), $A$ does not contain $z$ and
its neighbors. Also, each $z_i$ has $r-1$ neighbors in $A$. By
condition (b), each vertex in $A$ has at most $2$ neighbors in
$\{z_1,z_2,\ldots,z_s\}$. Then $|A| \ge s(r-1)/2$. Therefore,
$$
    |F_1 \cup F_2| \ge 1 +|N(z)| + |A| \ge 1 +r + s(r-1)/2.
$$
Also, by the choice of $z$, each node $z_i$ has at most $s$
neighbors in $\symdiff(F_1,F_2)$ and hence at least $r-s$ vertices
in $F_1 \cap F_2$, which are not neighbors of $z$.  This further
implies that $|F_1 \cap F_2| \ge 2(r-s)$.  Then,
$$
   4r-4 \ge |F_1| + |F_2|
         =  |F_1 \cup F_2| + |F_1 \cap F_2|
         \ge (1+r+s(r-1)/2) +2(r-s) = 3r+1+s(r-5)/2.
$$
As $r\ge 5$ and $s\ge 2$, this inequality in fact is an equality
and also $r=5$ or $s=2$. It is also the case that $|(F_1\cap
F_2)-N(z)|=r-s$, and each $z_i$ is adjacent to any vertex in
$(F_1\cap F_2)-N(z)$. That is, $(F_1\cap F_2)\cap A=(F_1\cap
F_2)-N(z)$.

If $2 \le s \le r-3$, then $|(F_1\cap F_2)-N(z)| \ge 3$ and so
$|N(z_1)\cap N(z_2)| \ge 3$, contradicting condition (b). If $r=5$
and $2=r-3<s\le 4$, then $|(F_1\cap F_2)\cap A|=|(F_1\cap
F_2)-N(z)|\ge 1$ and so $|N(z)\cap N(a)| \ge s
> 2$ for any $a\in (F_1\cap F_2)\cap A$, again impossible. Therefore, $r=s=5$ and
$F_1\cap F_2=\emptyset$. In this case, $A$ has $10$ vertices each
adjacent to exactly two vertices in $N(z)$.  Also, by condition
(b), two distinct vertices in $A$ have distinct pair of neighbors
in $N(z)$. For $I=\{i,j\}$, we can use $z_I$ to name the vertex of
$A$ adjacent to $z_i$ and $z_j$.  By condition (a), we also have
that $z_I$ is not adjacent to those $z_J$ with $I\cap J \ne
\emptyset$ and hence adjacent to those $z_K$ with $I\cap K
=\emptyset$. So, $G$ is in fact $G_5$, a contradiction. \qed

\bigskip

According to Lemma \ref{Lemma 8} and the definition of
diagnosability of a system under the PMC model using the
pessimistic diagnosis strategy, we have

%***   Theorem 9  The PMC model under pessimistic diagnosis strategy******
\begin{thm} \label{Theorem 9}
If $r \ge 5$ and $G$ is an $r$-regular graph, which is not
isomorphic to $G_5$, then $G$ is $(2r-2)/(2r-2)$-diagnosable under
the PMC model using the pessimistic strategy if the following two
conditions hold.

{\rm(a)} $G$ is triangle-free.

{\rm(b)} $|N(u) \cap N(v)| \le 2$ for every two distinct
         nodes $u$ and $v$ of $G$.
\end{thm}

\proof Suppose to the contrary that $G$ is not
$(2r-2)/(2r-2)$-diagnosable. Then, by Definition \ref{def 4},
there exist two indistinguishable and hence distinct sets $F_1$
and $F_2$ with $|F_1| \le 2r-2$ and $|F_2| \le 2r-2$ but $|F_1
\cup F_2| > 2r-2$. According to Lemma \ref{Lemma 8}, there exists
a node $w \in \symdiff(F_1,F_2)$ adjacent to some $x\not\in F_1
\cup F_2$. Without loss of generality, we may assume that $w \in
F_1 -F_2$. Choose a syndrome $\sigma \in \Omega(F_1) \cap
\Omega(F_2)$. If $\sigma(x, w)=0$ (respectively, $\sigma(x,w)=1$),
then $F_1$ (respectively, $F_2$) is not an allowable fault set
with respect to $\sigma$, a contradiction. \qed

\bigskip

Next, we establish  diagnosabilities for regular networks under
MM* model using the precise diagnosis strategy.

%*********     The MM* model under pessimistic diagnosis strategy****
\begin{thm} \label{Theorem 10}
 If $r \ge 6$ and $G=(V,E)$ is an $r$-regular graph,
 then $G$ is $(2r-2)/(2r-2)$-diagnosable under the MM* model
 using the pessimistic strategy
 if the following two conditions hold.

 {\rm (a)} $G$ is triangle-free.

 {\rm (b)} $|N(u) \cap N(v)| \le 2$ for every two
           distinct nodes $u$ and $v$ of $G$.
\end{thm}
\proof Suppose to the contrary that $G$ is not
$(2r-2)/(2r-2)$-diagnosable. Then, by Definition \ref{def 4},
there exist two indistinguishable and hence distinct sets $F_1$
and $F_2$ with $|F_1| \le 2r-2$ and $|F_2| \le 2r-2$ but $|F_1
\cup F_2| > 2r-2$. According to Lemma \ref{Lemma 8},
$\symdiff(F_1,F_2)$ has at least one node $w$ adjacent to some
node $x \not\in F_1 \cup F_2$. Denote $F_3$ the set of all such
nodes $x$.  Since $F_1$ and $F_2$ are indistinguishable, none of
the conditions in Lemma \ref{Lemma 4} holds.  It follows that for
any node $v \in F_3$,

\medskip
 (i) $|N(v)\cap(F_1-F_2)|\leq 1$,

(ii) $|N(v)\cap(F_2-F_1)|\leq 1$,

(iii) $N(v) \subseteq F_1 \cup F_2$.

\medskip \noindent
Then, $F_3$ is an independent set with $N(F_3) \subseteq F_1 \cup
F_2$. Also, (i) and (ii) and $|N(v)|=r$ imply $|N(v) \cap F_1 \cap
F_2| \ge r-2$, which gives $|F_1 \cap F_2| \ge r-2$ and so
$$
    |F_1 \cup F_2| = |F_1| +
    |F_2| - |F_1 \cap F_2| \le (2r-2) + (2r-2) -(r-2) = 3r-2.
$$

We first claim that $|N(w) \cap F_3| \le 2$ for each node $w \in
\symdiff(F_1,F_2)$. Assume to the contrary that
$\symdiff(F_1,F_2)$ has a node $w$ adjacent to three distinct
nodes $p_1$, $p_2$ and $p_3$ in $F_3$. Then, (i) to (iii) imply
$|N(p_i) \cap (F_1 \cap F_2)| \ge r-2$ for $1 \le i \le 3$; and
condition (b) implies $|N(p_i) \cap N(p_j) \cap (F_1 \cap F_2)|
\le 1$ for $i \ne j$. Thus, $|F_1 \cap F_2| \ge (r-2)+(r-3)+(r-4)
=3r-9$, and so $|F_1\cup F_2| = |F_1|+|F_2|-|F_1 \cap F_2| \le
(2r-2) + (2r-2) - (3r-9) = r+5$. On the other hand, condition (b)
implies $|F_1 \cup F_2| \ge |N(p_1) \cup N(p_2) \cup N(p_3)| \ge
r+(r-2)+(r-4) \ge 3r-6>r+5$ as $r \ge 6$, a contradiction.

{\bf Case 1.} $|F_3| \ge 2$ and $|N(p) \cap (\symdiff(F_1,F_2))| =
1$ for each node $p \in F_3$.

Choose $p_1 \in F_3$ with $N(p_1) \cap(\symdiff(F_1,F_2)) =
\{w\}$. Also choose $p_2 \in (N(w) \cap F_3) - \{ p_1 \}$ if
$|N(w) \cap F_3|=2$, and any node $p_2 \in F_3- \{ p_1 \}$
otherwise.  By condition (b), $|F_1 \cap F_2| \ge |N(\{p_1, p_2
\}) \cap F_1 \cap F_2| \ge (r-1) + (r-3) = 2r-4$. So, $|F_1 \cup
F_2| = |F_1| + |F_2| - |F_1 \cap F_2|
                     \le (2r-2)+(2r-2)-(2r-4) = 2r$.

On the other hand, by condition (a), $N(w) \cap N(p_1)=\emptyset$.
If $w$ is (respectively, is not) adjacent to $p_2$, by condition
(a) (respectively, condition (b)),  $N(w) \cap N(p_2)=\emptyset$
(respectively, $|N(w) \cap N(p_2)| \le 2$). In either case,
$|N(w)-(N(\{ p_1, p_2 \} )\cup F_3)| \ge r-3$. Hence $|F_1 \cup
F_2| \ge |N(\{ p_1, p_2 \})| + |N(w)-(N(\{ p_1, p_2 \}) \cup F_3)|
                      \ge r+(r-2)+(r-3) > 2r$ as $r \ge 6$,
a contradiction to $|F_1 \cup F_2| \le 2r$.

{\bf Case 2.} $|F_3| \ge 2$ and $|N(p_1) \cap (\symdiff(F_1,F_2))|
\ge 2$ for some node $p_1 \in F_3$.

Assume that $p_1$ is adjacent to two distinct nodes $w_1$ and
$w_2$ in $\symdiff(F_1,F_2)$. Furthermore, assume that $|N(w_1)
\cap F_3| \ge |N(w_2) \cap F_3|$. Choose $p_2 \in (N(w_1) \cap
F_3) -\{ p_1 \}$ if $|N(w_1) \cap F_3|=2$, or  $p_2 \in (N(w_2)
\cap F_3) -\{ p_1 \}$ if $|N(w_2) \cap F_3|=2$, or $p_2 \in F_3
-\{ p_1 \}$ otherwise. By (i)--(iii) and condition (b), $|F_1 \cap
F_2| \ge |N(\{ p_1, p_2 \}) \cap (F_1 \cap F_2)| \ge (r-2) + (r-4)
= 2r-6$. Hence, $|F_1 \cup F_2| = |F_1| + |F_2| - |F_1 \cap F_2|
                    \le (2r-2) + (2r-2) - (2r-6) = 2r+2$.

\begin{figure}[htb]
\setlength{\unitlength}{0.2cm}%{0.15cm}
\begin{center}
\begin{picture}(35,23)(10,6)
%left
\put(00,10){\circle*{1.125}} \put(05,10){\circle*{1.125}}
\put(10,10){\circle*{1.125}} \put(15,10){\circle*{1.125}}
\put(20,10){\circle*{1.125}} \put(00,20){\circle*{1.125}}
\put(10,20){\circle*{1.125}} \put(20,20){\circle*{1.125}}

\put(-.5,8){$w_{1}$} \put(4.5,8){$x_{1}$} \put(9,8){$w_{2}$}
\put(14.5,8){$x_{2}$} \put(20,8){$x_{3}$} \put(-.5,21.5){$p_{1}$}
\put(9.5,21.5){$p_{3}$} \put(19.5,21.5){$p_{2}$}

\put(-1,25){$F_{3}$}\put(-1,5){$F_{1}\cup F_2$}

\put(00,20){\line(0,-1){10}} \put(00,20){\line(1,-1){ 10}}
\put(10,20){\line( 0,-1){10}} \put(20,20){\line( 0,-1){ 10}}
\put(20,20){\line( -2,-1){20}}\put(20,20){\line( -1,-2){5}}
\put(00,10){\line( 1,0){15}} \put(15,10){\oval(10,6)[b]}
\put(-2,19){\dashbox{0.2}(25,8)}\put(-2,4){\dashbox{0.2}(25,8)}
%right
\put(35,10){\circle*{1.125}} \put(40,10){\circle*{1.125}}
\put(45,10){\circle*{1.125}} \put(50,10){\circle*{1.125}}
\put(55,10){\circle*{1.125}} \put(35,20){\circle*{1.125}}
 \put(55,20){\circle*{1.125}}

\put(34.55,8){$w_{1}$} \put(39.5,8){$x_{1}$} \put(44,8){$w_{2}$}
\put(49.5,8){$x_{2}$} \put(55,8){$x_{3}$}
\put(34.55,21.5){$p_{1}$} \put(54.55,21.5){$p_{2}$}

\put(34,25){$F_{3}$}\put(34,5){$F_{1}\cup F_2$}

\put(35,20){\line(0,-1){10}} \put(35,20){\line(1,-1){ 10}}
\put(55,20){\line( 0,-1){ 10}} \put(55,20){\line( -1,-2){5}}
\put(35,10){\line( 1,0){15}} \put(50,10){\oval(10,6)[b]}
\put(33,19){\dashbox{0.2}(25,8)}\put(33,4){\dashbox{0.2}(25,8)}
\end{picture}
\end{center}
\caption{Relation among $N(w_1)$, $N(w_2)$, $N(\{p_1,p_2\})$ and
$F_3$.} \label{Figure 6}
\end{figure}
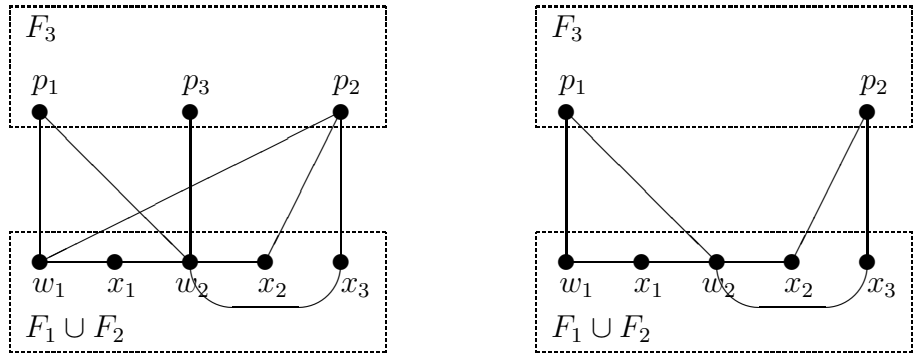

On the other hand, condition (b) assures $|N(\{ p_1, p_2 \})| \ge
r + (r-2) = 2r-2$. If $w_1$ is adjacent to $p_2$, then by
conditions (a) and (b), $|N(w_1) - (N(\{ p_1, p_2 \}) \cup F_3)|
\ge r-2$ and $|N(w_2) - (N(w_1) \cup N(\{ p_1, p_2 \}) \cup F_3)|
\ge r-5$. See the left of Figure \ref{Figure 6}. Similarly, if
$w_1$ is not adjacent to $p_2$ (as $N(w_1) \cap F_3=1$ ), then
$|N(w_1) - (N(\{ p_1, p_2 \}) \cup F_3)| \ge r-3$ and $|N(w_2) -
(N(w_1) \cup N(\{ p_1, p_2 \}) \cup F_3)| \ge r-4$. See the right
of Figure \ref{Figure 6}. It follows that $|N(w_1) \cup
N(w_2)-(N(\{ p_1, p_2 \}) \cup F_3)| \ge 2r-7$. Hence, $|F_1 \cup
F_2| \ge |N(\{ p_1, p_2 \})| + |N(w_1) \cup N(w_2)-(N(\{ p_1, p_2
\}) \cup F_3)| \ge 4r-9 \ge 2r+3$ as $r \ge 6$, a contradiction to
$|F_1 \cup F_2| \le 2r+2$.

{\bf Case 3.} $|F_3|=1$, say $F_3 = \{p\}$.

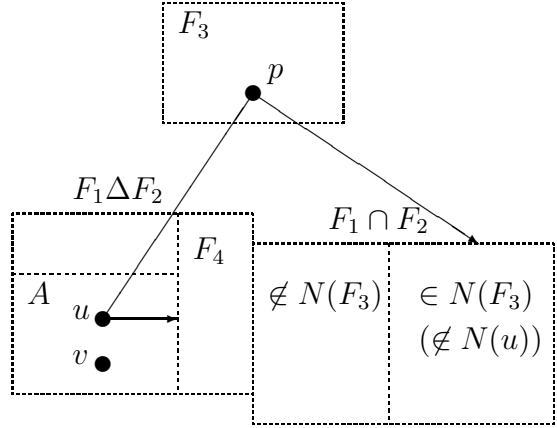
\begin{figure}[htb]
\setlength{\unitlength}{0.2cm}%{0.15cm}
\begin{center}
\begin{picture}(27,25)(00,8)
%left
 \put(16,27){\circle*{1.125}}
 \put(06,9){\circle*{1.125}}
 \put(06,12){\circle*{1.125}}

 \put(11,31){$F_3$}
 \put(17,28){$p$}
 \put(4,9){$v$}
 \put(4,12){$u$}
 \put(1,13){$A$}
 \put(12,16){$F_4$}
 \put(4,20){$F_1\Delta F_2$}
 \put(21,18){$F_1\cap F_2$}

 \put(17,13){$\not\in N(F_3)$}
 \put(27,13){$\in N(F_3)$}
 \put(27,10){$(\not\in N(u))$}
 \put(16,27){\vector(3,-2){15}}
 \put(06,12){\vector(1,0){5}}

 \put(6,12){\line(2,3){10}}
 \put(0,7){\dashbox{0.2}(16,12)}
 \put(16,5){\dashbox{0.2}(20,12)}
 \put(10,25){\dashbox{0.2}(12,8)}
 \multiput(11,19)(0,-0.5){24}{\line(0,-1){0.2}}
%\multiput(16,19)(0,-0.5){24}{\line(0,-1){0.2}}
 \multiput(25,17)(0,-0.5){24}{\line(0,-1){0.2}}
 \multiput(00,15)(0.5,0){22}{\line(1,0){0.2}}

\end{picture}
\end{center}
\caption{Relation among $N(u)$, $N(F_3)\cap F_1\cap F_2$ and $F_4$
for Case 3. The two arrows represent that $p$ and $u$ are adjacent
to all nodes in $N(F_3)\cap F_1\cap F_2$ and $F_4$, respectively.}
\label{Figure 7}
\end{figure}

Since $|\symdiff(F_1,F_2)| \ge 2$, we may choose two nodes $u$ and
$v$ from $\symdiff(F_1,F_2)$ such that $p \in N(u)$. Let $A = \{u,
v\}$. Notice that $A \cup (N(A)-F_3) \cup N(F_3) \subseteq F_1
\cup F_2$.   By conditions (a) and (b), $|A \cup (N(A)-F_3)| \ge
2r-1$ and $|N(F_3) - (A \cup N(A))| \ge r-3$. Then, $|F_1\cup F_2|
\ge (2r-1) + (r-3) = 3r-4$. We assume that $|F_1 \cup F_2| =
3r-4+s$, where $s \ge 0$. Hence, $|F_1 \cap F_2| = |F_1| + |F_2| -
|F_1 \cup F_2|
              \le (2r-2) + (2r-2) - (3r-4+s) = r-s$.
Also, $ |N(F_3) \cap (F_1 \cap F_2)| \ge r-2 $. By condition (a),
$|N(u) \cap (F_1\Delta F_2)| \ge |N(u)|- (|F_1 \cap F_2| -|N(F_3)
\cap (F_1 \cap F_2)|) -|F_3|\ge r-((r-s)-(r-2))-1=r-3+s$. Refer to
Figure \ref{Figure 7}. Let $F_4=N(u) \cap (\symdiff(F_1,F_2))$ and
$\alpha=|(N(F_4)-(A \cup N(A))) \cup N(F_3)|$. By conditions (a)
and (b), for each node $x \in F_4$, $|N(x) - (A \cup N(A))| \ge
r-3$. Then, by condition (b), $\alpha \ge (|F_4|(r-3) - |N(F_3) -
A|)/2
       \ge ((r-3+s)(r-3)-(r-1))/2 \ge 2+3s/2$ as $r \ge 6$.
Hence, $|F_1 \cup F_2| \ge |A \cup (N(A)-F_3) \cup N(F_3)| +
\alpha
                \ge (2r-1)+(r-3)+(2+3s/2) =3r-2+3s/2$,
a contradiction to $|F_1 \cup F_2| = 3r - 4 + s$. \qed

%************************************************************************
%                                Section 4
%************************************************************************
%\section{Diagnosabilities of multiprocessor systems}
\section{Application to multiprocessor systems using regular networks}

In this section we apply the four theorems in Section 3 to eight
popular multiprocessor systems, while it is also possible to apply
them to many other potentially useful ones not shown here. To
introduce these systems, we need the following notations. Define
$[m]=\{0,1,\ldots,m-1\}$ and $
     [m]^n = \{x_{n-1} x_{n-2} \ldots  x_0: x_i \in [m]
      \mbox{ for }  i \in [n]\},
$ where $m$ and $n$ are positive integers. Let
$x=x_{n-1}x_{n-2}\ldots x_0 \in [m]^n$ and $y=y_{n-1}y_{n-2}\ldots
y_0 \in [m]^n$. The {\it Hamming distance} of $x$ and $y$, denoted
by $H(x,y)$, is the number of indices $i$ such that $x_i\ne y_i$.

%******************************** Example 1

\begin{ex}
          {\bf  Hypercube} {\boldmath $Q_n$} {\bf \cite{c30}}
\end{ex}

A {\it hypercube}  of $n$ dimensions can be expressed by a graph
$Q_n=(V,E)$ with $V=[2]^n$ and $E=\{(x,y): H(x,y)=1\}$.

%******************************** Example 2
\bigskip
\begin{ex}
          {\bf  Enhanced hypercube} {\boldmath
          $EQ_{n,s}$} {\bf \cite{c31}}
\end{ex}

An enhanced hypercube is just a hypercube augmented with certain
extra
links. % called {\it skips}.
More precisely,  an {\it $(n,s)$-enhanced hypercube}
 can be expressed by a graph $EQ_{n,s}=(V,E)$ with $V=[2]^n$ and $
    E = \{(x,y) : H(x,y)=1 \mbox{ or } y=x_{n-1}x_{n-2} \ldots
                        x_{s+1}\bar{x}_s \bar{x}_{s-1}\ldots
                        \bar{x}_0$ for some $0 \le s \le n-1 \}$,
where  $\bar{x}_i = 1-x_i$ for $0 \le i \le s$.

%******************************* Example 3

\begin{ex}
          {\bf Twisted cube} {\boldmath $TQ_n$ }{\bf \cite{c11}}
\end{ex}

Assume that $n$ is odd. Define $P_j(x) = (x_j+x_{j-1}+\ldots+x_0)$
mod 2, where $0 \le j \le n-1$.  A {\it twisted cube} of $n$
dimensions can be expressed by a graph $TQ_n=(V,E)$ with $V=[2]^n$
and $E$ consisting of all $(x,y)$'s that satisfy the following two
conditions for some $0 \le k \le (n-1)/2$:

\noindent (1) $x_{2k}x_{2k-1}=\bar{y}_{2k}y_{2k-1}$ or
($x_{2k}x_{2k-1}=y_{2k}\bar{y}_{2k-1}$ and $P_{2k-2}(x)=1$) or
($x_{2k}x_{2k-1}=\bar{y}_{2k}\bar{y}_{2k-1}$

\noindent ~~~~ and $P_{2k-2}(x)=0$);

\noindent (2) $x_{2j}x_{2j-1}=y_{2j}y_{2j-1}$ for all $j \ne k$,

\noindent where $x_0x_{-1}$ is regarded as $x_0$ when $k=0$.

%******************************* Example 4

\begin{ex}
          {\bf M\"{o}bius cube} {\boldmath $MQ_n$}{\bf \cite{c29}}
\end{ex}

A {\it M\"{o}bius cube} of $n$ dimensions can be expressed by a
graph $MQ_n=(V,E)$ with $V=[2]^n$ and $E$ containing those
$(x,y)$'s with
 $y= x_{n-1} x_{n-2} \ldots x_{i+2} \,0  \, \bar{x}_i x_{i-1}
\ldots x_0$ \,  or \, $y=x_{n-1} x_{n-2} \ldots x_{i+2} \, 1 \,
\bar{x}_i \bar{x}_{i-1} \ldots\bar{x}_0$ for some $0 \le i \le
n-2$. Besides, $E$ contains ($x$, $\bar{x}_{n-1} x_{n-2} \ldots
x_0$) or ($x$, $\bar{x}_{n-1}\bar{x}_{n-2} \ldots \bar{x}_0$) but
not both.

%******************************** example 5

\begin{ex}
          {\bf Crossed cube} {\boldmath $CQ_n$ }{\bf \cite{c28}}
\end{ex}

A {\it crossed cube} of $n$ dimensions can be expressed by a graph
$CQ_n = (V,E)$ with $V=[2]^n$ and $E$ consisting of all $(x, y)$'s
that satisfy the following conditions for some $1 \le m \le n$:

\noindent (1) $x_{n-1} x_{n-2} \ldots x_{m}x_{m-1} = y_{n-1}
y_{n-2} \ldots y_{m}\bar{y}_{m-1}$;

\noindent (2) $x_{m-2}=y_{m-2}$ if $m$ is even;

\noindent (3) $(x_{2i+1}x_{2i},y_{2i+1}y_{2i}) \in \{(00,00),
(10,10), (01,11), (11,01)\}$
   for $0 \le i \le \lfloor (m-1)/2  \rfloor-1$.

%******************************** Example 6

\begin{ex}
          {\bf Cube-connected cycles} {\boldmath $CCC_n$}
          {\bf \cite{c33}}
\end{ex}

Cube-connected cycles can be obtained by replacing each node of a
hypercube with a cycle. More precisely,  {\it cube-connected
cycles} of $n$ dimensions can be expressed by a graph
$CCC_n=(V,E)$ with $V=\{[x, i]: x\in [2]^n \mbox{ and } i\in
[n]\}$ and $
     E = \{([x, i], [x, j]): x \in [2]^n, \, i,j \in [n]  \mbox{ and } j\equiv(i \pm 1) \bmod n\}
     \cup\{([x, i], [y, i]): x, y \in [2]^n, \,  i \in [n]  \mbox{ and}$
     $y= x_{n-1} x_{n-2} \ldots x_{i+1} \bar{x_i} x_{i-1}
\ldots x_0 \}$.

%******************************** Example 7

\begin{ex}
          {\bf Torus} {\boldmath $T_n(m)$} {\bf \cite{c34}}
\end{ex}

An {\it  $m$-sided torus} of $n$ dimensions can be expressed by a
graph $T_n(m)=(V,E)$ with
     $V=[m]^n$ and
$
    E = \{(x,y):  y_i \equiv (x_i\pm 1) \bmod m \mbox{ for some $i \in [n]$ and }
                  x_j = y_j \mbox{ for all $j\ne i$}\}.
$

%***************************** Example 8

\begin{ex}
          {\bf Star graph} {\boldmath $S_n$} {\bf \cite{c32}}
\end{ex}

A {\it star graph} of $n$ dimensions can be expressed by a graph
$S_n=(V,E)$ with $V$ being the set of all permutations of $\{1, 2,
\ldots, n\}$, and $E$ consisting of all $(u, v)$'s such that  $u =
u_1 u_2 \ldots u_k \ldots u_n$ and $v = u_k u_2 \ldots
u_{k-1}u_1u_{k+1} \ldots u_n$ (i.e., swap $u_1$ and $u_k$) for
some $2 \le k \leq n$.

\bigskip

The diagnosabilities of these multiprocessor systems can be
determined by the aid of Theorems \ref{Theorem 6}, \ref{Theorem
7}, \ref{Theorem 9} and \ref{Theorem 10}. We first have to check
if they satisfy the conditions in these theorems. As the checking
is easy, we only summarize the  results in Table I. Consequently,
we have their diagnosabilities, as shown in Table II.

\bigskip
%\newpage
\begin{center}
              Table I: Properties of multiprocessor systems.
\end{center}
{\small

 \noindent
\begin{tabular}{|c|c|c|c|c|c|c|c|} \hline

  system               & $r$-regular             & triangle-free    &
  $G_{r+1,r+1}$        & $G_8$                   & $G_5$            &
  $N(u) \neq N(v)$     & $|N(u)\cap N(v)| \le 2$ \\ \hline \hline

  $Q_n$                & $r=n$                   & yes              &
  $\not\cong$          & $\not\cong$             &  $\not\cong$     &
  yes if $n\ge 3$      & yes if $n\ge 2$         \\ \hline

  $EQ_{n,s}$           & $r=n+1$                 & yes if $s\ge 2$  &
  $\not\cong$          & $\not\cong$             & $\not\cong$      &
  yes if $n\ge 3$      &  yes if $n\ge 2,s \ne 2$   \\ \hline

  $TQ_n$               & $r=n$                   & yes              &
  $\not\cong$          & $\not\cong$ if $n\ne 3$ & $\not\cong$      &
  yes if $n\ge 3$      & yes if $n\ge 2$         \\ \hline

  $CQ_n$               & $r=n$                   & yes              &
  $\not\cong$          & $\not\cong$ if $n\ne 3$ & $\not\cong$      &
  yes if $n\ge 3$      & yes if $n\ge 2$         \\ \hline

  $MQ_n$               & $r=n$                   & yes              &
  $\not\cong$          & $\not\cong$ if $n\ne 3$ & $\not\cong$      &
  yes if $n\ge 3$      & yes if $n\ge 2$         \\ \hline

  $CCC_n$              & $r=3$ if $n \ge 3$      & yes if $n \ne 3$   &
  $\not\cong$          & $\not\cong$             & $\not\cong$      &
  yes                  & yes                     \\ \hline

  $T_n(m)$             & $r=2n$                  & yes if $m \ne 3$ &
  $\not\cong$          & $\not\cong$             & $\not\cong$      &
  yes if $n\ge 3$      & yes if $n\ge 2$         \\ \hline

  $S_n$                & $r=n-1$                 & yes              &
  $\not\cong$          & $\not\cong$             & $\not\cong$      &
  yes                  & yes                     \\ \hline

\end{tabular}
}

{\footnotesize %\noindent $\cong$: isomorphic.
               \noindent $\not\cong$: not isomorphic.

               %\noindent $T_n(m)$: consider only for $m>4$.
               \noindent $N(u) \ne N(v)$: $N(u) \neq N(v)$ for any two
                                          distinct nodes $u$ and $v$ in
                                          $V$.

               \noindent $|N(u) \cap N(v)| \le 2$:
               $|N(u) \cap N(v)| \le 2$ for any two distinct
               nodes $u$ and $v$ in $V$.

               }

\bigskip

\begin{center}
              Table II: Diagnosabilities of multiprocessor
              systems.
\bigskip

\noindent
\begin{tabular}{|l|c|c|c|c|} \hline

system                 & \multicolumn{2}{c|}{PMC}
                       & \multicolumn{2}{c|}{MM*} \\ \cline{2-5}

                       & precise                 & pessimistic
                       & precise                 & pessimistic \\ \hline \hline

$Q_n$                  & $n$ {\bf \cite{c14}}    & $2n-2/2n-2$
{\bf \cite{c6}}
                       & $n$ {\bf \cite{c18}}    & $2n-2/2n-2$ \\ \hline

$EQ_{n,s}$             & $n+1$ {\bf \cite{c16}}  & $2n/2n$ {\bf
\cite{c16}}
                       & $n+1$ {\bf \cite{c18}}  & $2n/2n$ \\ \hline

$TQ_n$                 & $n$                     & $2n-2/2n-2$
                       & $n$                     & $2n-2/2n-2$ \\ \hline

$CQ_n$                 & $n$                     & $2n-2/2n-2$
                       & $n$  {\bf \cite{c23}}   & $2n-2/2n-2$ \\ \hline

$MQ_n$                 & $n$ {\bf \cite{c17}}    & $2n-2/2n-2$
{\bf \cite{c17}}
                       & $n$                     & $2n-2/2n-2$ \\ \hline

$CCC_n$                & $n+2$                   & $2n+2/2n+2$
                       & $n+2$                   & $2n+2/2n+2$ \\ \hline

$T_n(m)$               & $2n$                    & $4n-2/4n-2$
                       & $2n$                    & $4n-2/4n-2$ \\ \hline

$S_n$                  & $n-1$ {\bf \cite{c25}}  & $2n-4/2n-4$
{\bf \cite{c25}}
                       & $n-1$                   & $2n-4/2n-4$ \\ \hline

\end{tabular}
\end{center}

%\indent\indent
\hskip 0.7cm{\footnotesize
 [$i$]: also obtained in [$i$]; all others are results of this paper.
 }

%************************************************************************
%                                Section 5
%************************************************************************
\section{Conclusion}

Fault diagnosis of multiprocessor systems has received much
attention since Preparata {\it et al}. \cite{c2} introduced the
concepts of one-step diagnosis and sequential diagnosis. The {\it
one-step diagnosis} requires that all faulty nodes are found out
by decoding the syndrome, whereas the {\it sequential diagnosis}
consists of several diagnosis and repair phases. In each phase,
one or more faulty nodes will be determined and then repaired. The
process is iterated until all faulty nodes are repaired.

The {\it one-step diagnosability} of a multiprocessor system $S$
was defined to be the maximum number of faulty nodes allowed in
$S$ such that the one-step diagnosis of $S$ can be performed. The
{\it sequential diagnosability} of $S$ was defined similarly. In
\cite{c26}, the problem of computing the sequential diagnosability
for a general system was proved co-NP complete. In \cite{c35},
lower bounds on sequential diagnosabilities of grids and
hypercubes were suggested.

In \cite{c39}, Maheshwari and Hakimi introduced a probabilistic
model for fault diagnosis. A $p$-{\it probabilistically
diagnosable} system requires that any set of faulty processors
having a priori probability greater than or equal to $p$ of
occurring is uniquely diagnosable. In \cite{c42}, the problem of
determining whether a general system is $p$-probabilistically
diagnosable or not was proved co-NP complete.  A method of
achieving an optimal diagnosis with maximum probability was
presented in \cite{c40}.
%The main difficulty focuses on minimizing the number of tests and
%guaranteeing a high probability of correct diagnosis \cite{c40},
%\cite{c41}, \cite{c42}.
%which can correctly diagnose the state of every processor in the system .
In \cite{c36}, a probabilistic diagnosis algorithm was proposed
whose probability of correct diagnosis could approach one if a
slightly greater than linear number of tests were performed.

Another probabilistic diagnosis algorithm was proposed and
evaluated in \cite{c38}, on the basis of the concept that an
aggregate of maximum cardinality is fault-free with probability
approaching one if the cardinality of the actual fault set is
smaller than the syndrome-dependent diagnosability. The {\it
syndrome-dependent diagnosability} of a multiprocessor system is
determined by evaluating the cardinality of the smallest
consistent fault set that contains an aggregate of maximum
cardinality. Lower bounds on syndrome-dependent diagnosabilities
of toroidal grids and hypercubes were derived in \cite{c37}.

 In this paper, we have successfully
computed one-step diagnosabilities of eight regular multiprocessor
systems for two diagnosis models (i.e., the PMC and comparison
models) and two diagnosis strategies (i.e., the precise and
pessimistic diagnosis strategies). Our results were obtained as a
consequence of four sufficient conditions. Compared with most of
previous works which computed diagnosabilities only for individual
systems, the four sufficient conditions can derive
diagnosabilities for a class of regular systems. Our further
research interests include computing sequential diagnosabilities
and syndrome-dependent diagnosabilities of various systems for
different diagnosis models and
 diagnosis strategies.

%%%%%%%%%%%%%%%%%%%%%%%
\bigskip

\noindent {\bf Acknowledgements.} The authors thank the referees
for many constructive suggestions which make the paper much more
readable.

%************************************************************************
%                                Reference
%************************************************************************

\end{document}